\documentstyle[12pt]{article}
\begin{document}
\voffset 1cm
\setlength{\parindent}{5ex}
\newcommand{\be}{\begin{equation}}
\newcommand{\ee}{\end{equation}}
\newcommand{\ba}{\begin{eqnarray}}
\newcommand{\ea}{\end{eqnarray}}
\newcommand{\lb}{\label}
\newcommand{\bb}{\bibitem}
\newcommand{\dd}{\partial}
\newcommand{\half}{\frac{1}{2}}
\newcommand{\nn}{\nonumber}

\begin{center}
{\Large {\bf {
On the gravitational field of static and stationary 
 axial symmetric  bodies with multi-polar structure}}}
 
\vspace*{0.9cm}
{\large Patricio S. Letelier\footnote{e-mail: 
letelier@ime.unicamp.br} } 

\vspace{0.6cm}

{\em Departamento de Matem\'atica Aplicada-IMECC,
Universidade Estadual de Campinas
13081-970 Campinas. S.P., Brazil }

\end{center}
\vspace{0.4cm}
\baselineskip 0.6cm  

\centerline{ \small \bf Abstract}
\vspace{0.5cm}

We  give a physical interpretation to the multi-polar Erez-Rozen-Quevedo
 solution of the Einstein Equations  in terms of bars.  We find that each
multi-pole correspond to the Newtonian potential of a bar with linear density
proportional to a Legendre Polynomial. We use this fact to find an integral
 representation of the $\gamma$ function.  These integral representations are 
used in the context of the inverse scattering method  to find solutions 
associated to  one or more rotating bodies each one with their own
 multi-polar structure.


\newpage
\baselineskip 0.6cm 
\section{INTRODUCTION}

The adequate description of the gravitation field of an astrophysical
object has been an important subject in both relativistic and Newtonian 
gravity since their origin. The  particular case of the gravity associated 
to an axially symmetric body has played a central role in this theme. 
In Newtonian theory the gravitational potential of an axially symmetric body
can be always represented by its usual expansion in terms of Legendre
 polynomials (zonal harmonics). In general relativity we have that the Einstein
equations for an axially symmetric spacetime \cite{weyl},
\be
ds^2=-e^{2\psi}dt^2 +e^{2\gamma-2\psi}(dr^2+dz^2)+r^2e^{-2\psi}d\varphi,
 \lb{m1}
\ee
reduces to
\ba
&&\hspace*{1cm}\psi_{,rr}+\psi_{,r}/r+\psi_{,zz}=0,\\ \lb{lapc} 
&&\gamma[\psi]=\int r[(\psi_{,r}^2-\psi_{,z}^2)dr+2\psi_{,r}\psi_{,z}dz],
\lb{gam}
\ea
where the functions $\psi$ and $\gamma$ depend only on $r$ and $z$.

The simplicity of the Weyl equation is rather deceiving since the potential
 $\psi$ that obeys the usual Newtonian Laplace equation has a different meaning
  in general relativity. The mono-polar solution of the Einstein equations  
(Schwarzschild's solution) is represented in these coordinates by a bar
 of constant linear density of length $2m$, i.e., a `potential' whose
 multi-polar expansions contains  all the multi-polar moments beyond the 
dipole. Newtonian images are useful 
in general relativity but need to be used carefully \cite{letolsol}. 
Erez and Rosen \cite{ER} found that in spheroidal coordinates 
\ba
x=(R_+ + R_-)/(2\sigma),\;\; y=(R_+ - R_-)/(2\sigma), \lb{pro}\\
 R_+=\sqrt{r^2+(z+\sigma)^2},\;\; R_-=\sqrt{r^2+(z-\sigma)^2}, \lb{Rpm}
\ea
with $\sigma=m$ and  $x\geq 1,\; -1\leq  y\leq 1$,
the potential $\psi$ outside an isolate axially symmetric body can be 
written as
\be
\psi=-\sum_{k=0}^{\infty}q_k Q_k(x) P_k(y),  \lb{psigen}
\ee
where $P_k$ and $Q_k$ are the Legendre polynomials and the Legendre
 functions of the second kind, respectively, and $q_k$ are constants. The
 mono-polar terms correspond
 exactly to the potential representative of the Schwarzschild
 bar ($q_0=1$, and $q_k=0$,  for $k\not =0$). The particular 
case $q_0=1, q_1=0, q_2=q$ and  $q_k=0$ for $k>2$ is the solution usually
 associated to Erez-Rosen and studied in text books \cite{books}.

The function $\gamma$ for the general solution (\ref{psigen}) is rather
 formidable and was first computed by Quevedo \cite{que1}. Also the physical 
meaning of the 
multi-polar moments was clarified in \cite{que1}. This is a rather important
point  that can be settled after the 
work of G\"ursel \cite{gursel} that shows the relation between the physically
 oriented definition 
of moments due to Thorne \cite{kip} and collaborators   and the more 
geometrically motivated definition of Geroch \cite{bob} and
 Hansen \cite{han} that are of simpler computation \cite{hoen}. Also the
moments for a charged stationary generalization of the Erez-Rosen-Quevedo
solution were studied \cite{que2}.

The aim of this paper is to give a physical interpretation to each term in the
multi-polar expansion (\ref{psigen}) in terms of  Newtonian potentials
 associated to bars like the one  of the above mentioned Schwarzschild bar. 
This interpretation is achieved in terms of  potentials associated with
 bars of length $2m$  with densities
proportional to Legendre polynomials.  This  integral representation 
of the function $\psi$ can be used to find a simple  integral representation
 of the non-trivial function $\gamma$.  Furthermore, an integral 
representation of the function $F$ used to generate stationary 
solutions in the context of the inverse scattering
 method (ISM) \cite{BZ} \cite{sol}  can be easily found.
An example of  solution generated by this method is a multi-polar
 deformed  Kerr-NUT solution. We also discuss a new solution of the
 Einstein equations that represents two or more Kerr-NUT solutions, each one
 with their own multi-polar deformations, located along an axis of symmetry.

 In  Sec. II, combining two identities involving Legendre functions we 
find the physical interpretation  of (\ref{psigen}) in terms of the potential
 associated to bars. We also study an integral representation of the $\gamma$
function. In Section. III, we give a summary of the main formulas of the ISM
and we discuss solitonic  solutions whose seeds   are a field of multi-poles 
($q_0=0,  q_k\not=0, k>0$). Particular cases are a ``deformed" Kerr solutions
 and  multiple ``deformed" Kerr solutions . We conclude in Sec. IV with a
 few remarks.

\section{BARS OF VARIABLE DENSITY AND THE $\gamma$ FUNCTION}

In this section we show the relation between the potential $\psi$ defined in 
(\ref{psigen}) and the Newtonian potential associated to bars.
  The first step is the  Jeffery identity \cite{WW} that relates Legendre
 functions
of second kind ($Q_n$) with Ferrers functions of the 
 first ($P^{n}_m$) and second kind ($Q^{m}_n$), also known  
as associated Legendre functions,
\be
\int_{\pi}^{\pi}Q_n (\frac{\bar{x}\cos t +\bar{y}\sin t + i \bar{z}}{c})
 \cos{mt} dt=
2\pi \frac{(n-m)!}{(n+m)!}Q^{m}_n(ir) P^{m}_n(\cos \vartheta) \cos(m\varphi)
 \lb{jeff}
\ee
with 
\be
 (\bar{x},\bar{y},\bar{z})= c[(\bar{r}^2+1)^{\half} \sin\vartheta\cos\varphi,
(\bar{r}^2+1)^{\half} \sin\vartheta\sin\varphi, \bar{r} \cos\theta]. \lb{jc}
\ee 
By doing $m=0$,  $i\bar{r}=u$, $\cos\vartheta=v$  in (\ref{jc}) and using 
 Neumann  formula,
\be
Q_n(\zeta)=\half \int_{-1}^{1}\frac{ P_n(t)} {\zeta -t}dt,   \lb{neu}
\ee
we find
\be 
2\pi P_n(v)Q_n(u)=\half \int_{-1}^{1} P_n(t')d t' \int_{-\pi}^{\pi}
 \frac{dt}{A+B\cos t + C\sin t},
\ee
with
\ba
A&=& uv-t', \nonumber\\
B&=&(1-u^2)^\half(1-v^2)^\half \cos\varphi ,\nonumber\\
C&=&(1-u^2)^\half(1-v^2)^\half \sin\varphi .\lb{ABC}
\ea
Hence
\be
Q_n(u)P_n(v)=\half\int_{-1}^{1}\frac{P_n(t')dt'}{\sqrt{(u^2-
1)(1-v^2)+(uv-t')^2}}. \lb{int1}
\ee
Note that this identity is valid for any complex variables $u$ and $v$. Now 
identifying  $(u,v)$ with  the spheroidal  coordinates $(x,y)$,, and 
using the fact that the inverse of (\ref{pro}) is:
\be
r=\sigma\sqrt{(x^2-1)(1-y^2)},  \;\;\;  z=\sigma xy , \lb{rz}
\ee
we find 
\be
Q_n(x)P_n(y)=\half\int_{-\sigma}^{\sigma}
\frac{P_n(z'/\sigma)dz'}{\sqrt{r^2+(z-z')^2}}.
\ee
Since the Newtonian potential of a bar of length $2\sigma$ with 
linear density $\lambda$  located symmetrically along the $z$-axis is
\be
\Phi_N= -\int_{-\sigma}^{\sigma}
\frac{\lambda(z')dz'}{\sqrt{r^2+(z-z')^2}}.
\ee
The  density associated to the potential $-Q_n(x)P_n(y)$ is
\be
\lambda_n(z)=\half P_n(z/\sigma), \lb{lam}
\ee
and the Newtonian mass,
\be
m_n=\half \int_{-\sigma}^{\sigma} P_n(z/\sigma) dz. \lb{mass}
\ee
Note that $m_n=0$ for  $n\not=0$ and  $m_0=\sigma=m$. In other
 words the mono-polar term is the only one that carries mass, fact 
that is not surprising.

A direct computation using the identities of \cite{letolsol} shows that
 the function $\gamma$ can be written as
\be
\gamma =\gamma_0 +\sum_{k=0}^{\infty} \sum_{l=0}^{\infty}q_l q_k
\int_{-\sigma}^{\sigma}d z'
   \int_{-\sigma}^{\sigma} d z'' P_k(z'/\sigma) P_l(z''/\sigma)
 \frac{r^2+(z-z')(z-z'')}{4(z'-z'')^2 R_{z'}R_{z''}},\lb{gam2}
\ee
where $\gamma_0$ is a constant and $R_{z'}=\sqrt{r^2+(z-z')^2}$, etc.
 This is a singular integral representation, the integration need to be taken
 as   the Cauchy principal part, i.e.,   $\lim_{\eta \rightarrow 0, \epsilon 
\rightarrow 0}  \int_{-\sigma+\eta}^{\sigma-\eta}\int_{-\sigma
+\epsilon}^{\sigma-\epsilon}$,
this allows us to extract the finite part of the integral, we are left with
 a constant of the type $\ln(0)$ that can be absorbed in the 
constant  $\gamma_0$.
In summary, the series  (\ref{psigen}) can be interpreted as the infinite sum 
of the potentials associated with bars of equal length, $2\sigma=2m$, and 
linear densities $\lambda_n(z)=P_n(z/\sigma)/2$. 

\section{STATIONARY SOLUTIONS}

 The vacuum Einstein equations for the stationary axial symmetric metric
\be 
ds^2=g_{ab}(r,z)dx^a dx^b+e^\nu (dr^2+dz^2), \label{m2}
\ee  
with $(x^a)=(x^0,x^1)=(t,\varphi)$,  and $det(g_{ab})=-r^2$ are  equivalent
to \cite{sol}, 
\ba 
&& \hspace {1.5cm}(rg_{,r} g^{-1})_{,r}+(rg_{,z} g^{-1})_{,z}=0, \lb{e1}\\
&&\nu=-\ln r -\frac{1}{4}\int r[tr(g_{,r}g^{-1}_{,r}-g_{,z}g^{-1}_{,z})dr+
2tr(g_{,r}g^{-1}_{,z})]dz, \lb{e2}
\ea
where $g \equiv(g_{ab})$ and $g^{-1}_{,r}\equiv (g^{-1})_{,r}$, etc.

Equation (\ref{e1}) is an 
integrable system of equations 
that is closely related to the principal sigma model
 \cite{cosgrove}. Techniques to actually find solutions of these equations 
 are B\"acklund 
transformations and the inverse scattering method, also 
 a third method  constructed with elements of the previous
 two is the ``vesture method", all these methods are 
closely related \cite{cosgrove}.
 The general metric that represents the nonlinear superposition of
several Kerr solutions with a Weyl solution  can be found by 
 using the inverse scattering  method  \cite{sol}. We find  the $N$ 
soliton solution
 \begin{eqnarray}
g_{ab}&=&r^{-N}\prod_{s=1}^N\mu_s (g^0_{ab}-\sum_{k,l=1}^4((\Gamma^{-
1})_{kl}N^{(k)}_a N^{l}_b/(\mu_k\mu_l))  \lb{sN}\\
g_{ab}^0&=&diag(- e^{2\psi},r^2e^{-2\psi}) \lb{seed}\\
\Gamma_{kl}&=&\frac{-\mu_k \mu_lC_0^{(k)}C_0^{(l)}e^{2(\psi-F_k -F_l)}+
 C_1^{(k)}C_1^{(l)}e^{-2(\psi-F_k-F_l)}r^2}{\mu_k \mu_l
 (r^2+\mu_k\mu_l)} \lb{Gam}\\
N_0^{(k)}&=&-C_0^{(k)}e^{2(\psi-F_k)},\hspace{0.1cm}N_1^{(k)}
=C_1^{(k)}(r^2/\mu_k)e^{-2(\psi-F_k)} \lb{Nk}\\
e^\nu&=& e^{\gamma-\psi}(c_N)^2 r^{-N^2/2}(\prod_{k=1}^N\mu_k 
)^{(N+1)}\prod_{ {\tiny\begin{array}{cc}
 k,l=1\\k>l\end{array}}}^N(\mu_k -\mu_l)^{-2} det\Gamma_{ab}  \lb{fN}\\
\mu_k&=&W_k-z+[(W_k-z)^2+r^2]^{1/2} \lb{mu}.
\end{eqnarray} 
One of most interesting case is the superposition of one black hole with 
a field of multi-poles, i.e., a two soliton solution (N=2). In this case
 the constants $C_0^{(k)}$ and $ C_1^{(k)}$ are related to the
 usual constants by \cite{BZ}
\begin{eqnarray} 
&& C_1^{1} C_0^{2}-C_0^{1} C_1^{2}=\sigma, \hspace{0.5cm}C_1^{1}
 C_0^{2}+C_0^{1} C_1^{2}=m \nn\\
 && C_1^{1} C_1^{2}-C_0^{1} C_0^{2}=-b, \hspace{0.5cm}C_1^{1} C_1^{2}
+C_0^{1} C_0^{2}=a \lb{Ck}
  \end{eqnarray}
And 
\begin{equation} 
\sigma^2=m^2+b^2-a^2. \lb{sigm}
\end{equation}
Also, the constants $W_k$ are
\be
 W_1=z_1+\sigma,\hspace{0.5cm}W_2=z_1-\sigma. \lb{Wk}
 \ee
The symbol $c_N$ that appears in $e^\nu$ is an arbitrary
 constant. $m,a, b$ and $z_1$
represent the mass, the angular momentum per unit of mass, 
the NUT parameter, and the position 
along the $z$-axis of the  black hole, respectively. 
 The functions $F_k$ are the solutions of the system of equations \cite{sol},
\ba
&&(r\dd_r -\lambda\dd_z+2\lambda\dd_\lambda)F=r\psi_r, \nn\\
&&(r\dd_z+\lambda\dd_r)F=r\psi_z, \lb{Flam}\\
&&F|_{\lambda=0}=\psi, \lb{ic}
\ea
evaluates along the poles  $\mu_k$, i.e., $F_k=F|_{\lambda=\mu_k}$, see
 also \cite{sol} for an integral representation of F.

In summary, the  N-soliton solution (\ref{sN})-(\ref{fN})(even N)
 represents the
superposition of a Weyl solution characterized by $(\psi,\gamma)$, with $ N/2$
rotating bars (Kerr-NUT solutions) located on $z=z_k$  with
 masses $m_k$, angular
 momenta per unit of mass $a_k$ and NUT charges $b_k$. These constants
are related by equations like (\ref{Ck}) and (\ref{sigm}). The case $N$ odd
was studied in \cite{sol}.

The only integration that is left is  the system of equations (\ref{Flam})
with the initial condition (\ref{ic}). It is easy to check that the function
\be
 F=-\half \sum_{k=1}^\infty q_k\int_{-\sigma}^{\sigma}
 \frac{P_k(z'/\sigma)(z'+
R_{z'})}{(z'+R_{z'}+\lambda)R_{z'} } dz' \lb{F}
\ee
satisfies  (\ref{Flam}) and (\ref{ic}). We have excluded the mono-polar term 
due to the fact that the massive bar will appear as a result of the
 application of the ISM. The last expression can be obtained from the
 fact that the solution to (\ref{Flam})-(\ref{ic}) associated with
 the particular 
solution of Laplace equation $1/R_{z'}$ is $(z+R_{z'})/(z+R_{z'}+\lambda)$.

Now let us comeback to  the two soliton solution. We have that the
 ISM produces 
a  Kerr-NUT bar of length $2\sigma$ located along the $z$-axis
 with its center on 
$z=z_1.$  Really the this bar is independent of the collection of bars 
that describes the multi-polar structure. To have a solution that represent
 an isolated body with a multi-polar structure we need to have bars of the
 same size located on the same position. This is achieved
 taking $z_1=0$ i.e., centering the  Kerr bar on the origin of the $z$-axis
 and instead of taking $\sigma=m$ as in the static solution we
 need $\sigma^2=m^2+b^2-a^2$. This solution  
is a special case of the one discussed in \cite{que2} that has electric charge.

The fact that the integral representation is written in 
 cylindrical coordinates  can be used to construct solutions representing 
two or more bodies each one with its independent multi-polar structure
 located along the $z$-axis in a simple way. In other words these coordinates
 allow us the separation of the different contributions without a considerable
 amount of  work. In principle these configurations
 are not gravitationally stable,  but the nonlinear interaction 
produces struts  and/or strings that keep the bodies apart, see for instance 
\cite{letolmem}.

To be more specific the metric functions associated to two rotating
 bodies with multi-polar structure can be obtained from applying
 the ISM to the Weyl solution
\be 
-\psi= \sum_{k=1}^\infty [q_{k}^{(1)}\int_{-\sigma_1}^{\sigma_1}
 \frac{P_k(z'/\sigma_1)}{R_{z'}^{(1)} } dz' +q_{k}^{(2)}\int_{-\sigma_2}^{
\sigma_2} \frac{P_k(z'/\sigma_2)}{R_{z'}^{(2)} } dz' ], \lb{psi2}
\ee
where
\be
 R_{z'}^{(k)}=\sqrt{r^2+(z-z_k -z')^2} .
\ee
Thus (\ref{psi2}) represents  two collections of massless  bars  each
 one formed by bars 
of equal size $2\sigma_1$ and  $2\sigma_2$  located
 along the $z$-axis around the points $z=z_1$ and $z=z_2$. The
 respective function $F$ is, 

\be
-2 F=\sum_{k=1}^\infty[ q_{k}^{(1)}\int_{-\sigma_1}^{\sigma_1}
 \frac{P_k(z'/\sigma)(z'+R_{z'}^{(1)})}{(z'+R_{z'}^{(1)}
+\lambda)R_{z'}^{(1)} } dz'
+  q_k^{(2)}\int_{-\sigma_2}^{\sigma_2}
 \frac{P_k(z'/\sigma)(z'+R_{z'}^{(2)})}{(z'
+R_{z'}^{(2)}+\lambda)R_{z'}^{(2)} } dz' ]\lb{F2}
\ee
To obtain the two Kerr bars we construct a four soliton solution $N=4$,
in this case we will have eight constants $C_{a}^k$ that are related to 
the usual constants by \cite{letolstrut}
\begin{eqnarray} 
&& C_1^{1} C_0^{2}-C_0^{1} C_1^{2}=\sigma_1, \hspace{0.5cm}C_1^{1}
 C_0^{2}+C_0^{1} C_1^{2}=m_1\\
 && C_1^{1} C_1^{2}-C_0^{1} C_0^{2}=-b_1, \hspace{0.5cm}C_1^{1} C_1^{2}
+C_0^{1} C_0^{2}=a_1\\
 && C_1^{3} C_0^{4}-C_0^{3} C_1^{4}=\sigma_2, \hspace{0.5cm}C_1^{3} 
C_0^{4}+C_0^{3} C_1^{4}=m_2\\
 && C_1^{3} C_1^{4}-C_0^{3} C_0^{4}=-b_2, \hspace{0.5cm}C_1^{3}
 C_1^{4}+C_0^{3} C_0^{4}=a_2 .
\end{eqnarray}
And 
\begin{equation} 
\sigma_1^2=m_1^2+b_1^2-a_1^2, \hspace{0.5cm} \sigma_2^2=m_2^2+b_2^2-a_2^2
\end{equation}
Also, the constants $W_k$ are
\begin{eqnarray}
 W_1=z_1+\sigma_1,\hspace{0.5cm}W_2=z_1-\sigma_1 \\
 W_3=z_2+\sigma_2, \hspace{0.5cm}W_4=z_2-\sigma_2
\end{eqnarray}
The generalization  for the case of $n=N/2$ isolated bodies each one 
with their own multi-polar structure is obvious. 

\section{DISCUSSION}

Multiple Kerr solutions with or without multi-polar structure
 has been discussed
 by several authors in different contexts and using a variety of
 solution generating techniques \cite{multi}. The four soliton representation 
of two Kerr solutions each one with their own multipolar deformations 
presented here has a clear
 meaning. Note that  the  multi-polar structure of the solution 
 is generated by 
the multi-poles of the Erez-Rosen-Quevedo 
solution that have a  precise meaning in  general relativity. 

For the single  deformed Kerr solution [solution (22)-(27) with $N=2$
 and papameters (28)-(30)] as well as for the static case  we did
 not study the elimination of conical singularities. The
 ISM like other similar
methods introduce singular transformations like $t\rightarrow t+ c \varphi$
via rotation of tetrads. 
This point together with the study of the horizons, the stability of
the two body configuration and its related  struts and rotating 
strings will be studied in a wider context. The study of struts
 and strings associated to  two Kerr metrics can be found
 in \cite{letolstrut}. Even though the strut are not physical objects (they
have negative energy density)  they can be used to extract useful information
like  the rate of emission of gravitational radiation of head-on
 collision of rotating black holes  \cite{ALS}.

When  we choose the constants $C_a^{k}$ in such a way tha the NUT parameters 
$b_k$ are zero  the ISM, for even N, generates asymptotically flat solutions
 as long as the seed solutions are also asymptotically flat \cite{sol}. This 
is the case for the solutions generated from (\ref{psigen}). Again in this
 analysis conical singularities are excluded.

\vspace{0.3cm}

\noindent
{\bf Acknowledgments}

I want to thank  CNPq and FAPESP for financial support. Also
 S.R.  Oliveira and W.M. Vieira for discussions.

\end{document}